\documentclass[letterpaper]{article} 
\usepackage{aaai25}
\usepackage{times}  
\usepackage{helvet}  
\usepackage{courier}  
\usepackage[hyphens]{url}  
\usepackage{graphicx} 
\usepackage{natbib}  
\usepackage{caption} 
\usepackage{booktabs}
\usepackage{algorithm}
\usepackage{algorithmic}
\usepackage{array}
\newcolumntype{H}{>{\setbox0=\hbox\bgroup}c<{\egroup}@{}}
\usepackage[table]{xcolor} 
\usepackage{colortbl}      

\usepackage{newfloat}
\usepackage{listings}
\DeclareCaptionStyle{ruled}{labelfont=normalfont,labelsep=colon,strut=off} 
\floatstyle{ruled}
\newfloat{listing}{tb}{lst}{}
\floatname{listing}{Listing}
\nocopyright
\title{FlashCommunication V2: Bit Splitting and Spike Reserving for Any Bit Communication}
\author{
    Qingyuan Li\textsuperscript{\rm 1}, 
    Bo Zhang\textsuperscript{\rm 1}, 
    Hui Kang\textsuperscript{\rm 2}, 
    Tianhao Xu\textsuperscript{\rm 2},
    Yulei Qian\textsuperscript{\rm 1}, 
    Yuchen Xie\textsuperscript{\rm 1}, 
    Lin Ma\textsuperscript{\rm 1}
}
\affiliations{
    \textsuperscript{\rm 1}Meituan\\
     \textsuperscript{\rm 2}NVIDIA\\

}

\usepackage{bibentry}

\begin{document}

\maketitle

\begin{abstract}
Nowadays, communication bottlenecks have emerged as a critical challenge in the distributed training and deployment of large language models (LLMs). This paper introduces FlashCommunication V2, a novel communication paradigm enabling efficient cross-GPU transmission at arbitrary bit widths. Its core innovations lie in the proposed \emph{bit splitting} and \emph{spike reserving} techniques, which address the challenges of low-bit quantization.
Bit splitting decomposes irregular bit widths into basic units, ensuring compatibility with hardware capabilities and thus enabling transmission at any bit width. Spike reserving, on the other hand, retains numerical outliers (i.e., minima and maxima) as floating-point numbers, which shrinks the dynamic numerical range and pushes the quantization limits to 2-bit with acceptable losses.
FlashCommunication V2 significantly enhances the flexibility and resource utilization of communication systems. Through meticulous software-hardware co-design, it delivers robust performance and reduced overhead across both NVLink-based and PCIe-based architectures, achieving a maximum \textbf{3.2$\times$} speedup in AllReduce and \textbf{2$\times$}in All2All communication.
\end{abstract}

\section{Introduction}


In the fast-evolving field of AI, large language models (LLMs) favor mixture-of-experts (MoE) for strong performance with efficient training and inference. DeepSeek-V3~\cite{deepseekai2025deepseekv3technicalreport}, a 671B-parameter mixture-of-experts (MoE) model with 37B activated parameters per token, requires only 2.788M H800 GPU hours for training. MiniMax-01~\cite{minimax2025minimax01scalingfoundationmodels} is a model with 32 experts and a total of 456 billion parameters (45.9 billion parameters activated) equipped with lightning attention, which allows a context window of several million tokens. Qwen3~\cite{yang2025qwen3technicalreport}'s flagship Qwen3-235B-A22B is also a competitive MoE model in coding, math, and general tasks, featuring hybrid thinking modes and support for 119 languages. Kimi K2~\cite{kimik2technicalreport}, a 1T-parameter MoE model (32B activated), focuses on frontier knowledge and agentic capabilities. Together, these models drive progress in performance, efficiency, and versatility, shaping AI's future across various domains.

\begin{figure}[h]
\centering
\includegraphics[width=\columnwidth]{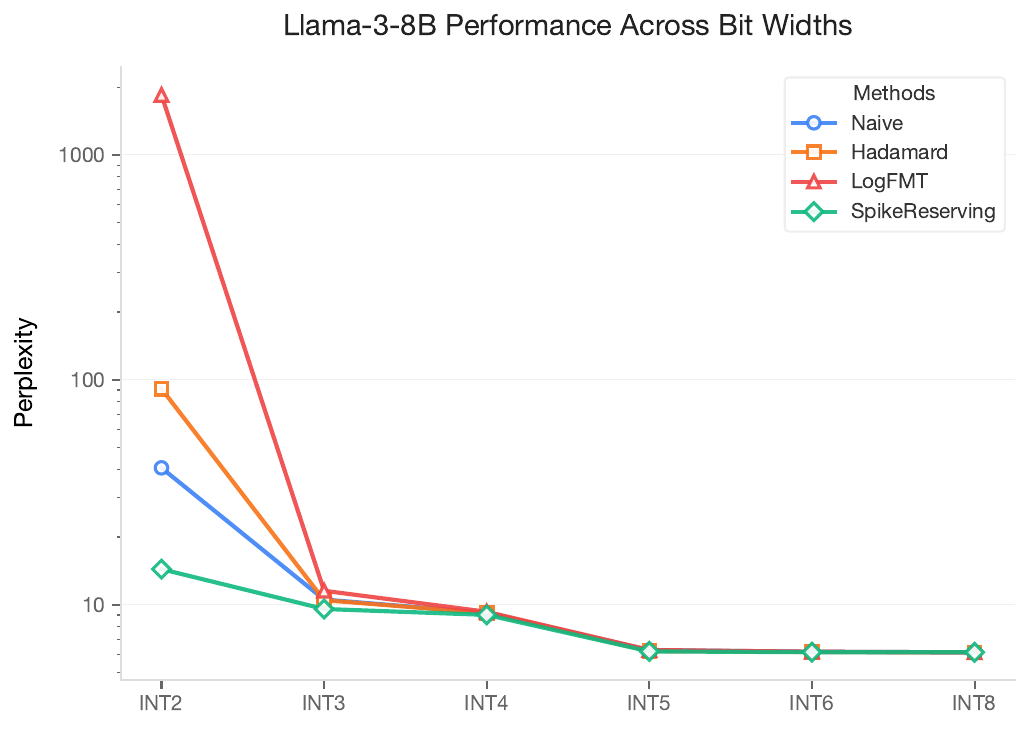} 
\caption{Llama-3-8B's C4 perplexity across common bit widths using various quantization schemes.}
\label{fig:c4-any-bit-width}
\end{figure}

To deploy trillion-parameter MoE LLMs in practice, it requires meticulous hardware-software codesign integrating multiple levels of parallelism, highly-optimized scheduling policies, and efficient kernel implementation. DeepSeek-V3 addresses hardware bottlenecks through innovations like Multi-head Latent Attention (MLA) for memory efficiency, Mixture of Experts (MoE) for computation-communication trade-offs, FP8 mixed-precision training, and a Multi-Plane Network Topology, while highlighting future hardware directions such as low-precision units and low-latency communication.

Huawei CloudMatrix384~\cite{zuo2025servinglargelanguagemodels} presents a next-gen AI datacenter architecture with 384 Ascend 910 NPUs and 192 Kunpeng CPUs interconnected via a high-bandwidth Unified Bus, paired with CloudMatrix-Infer for peer-to-peer serving, large-scale expert parallelism, and hardware-aware optimizations (e.g., INT8 quantization) to balance throughput and latency. Step-3~\cite{stepfun2025step3largeaffordablemodelsystem} achieves cost-effective decoding through Multi-Matrix Factorization Attention (MFA) to reduce KV cache and computation, and Attention-FFN Disaggregation (AFD) for distributed inference, outperforming DeepSeek-V3 in throughput. DeepEP~\cite{deepep2025,lmsys2025deepseek}, a communication library for MoE and expert parallelism, offers high-throughput/low-latency kernels (supporting FP8) optimized for prefill (normal dispatch) and decode (low-latency dispatch) phases, with features like SM-free communication-computation overlap. 



Prefill-decoding disaggregation has also become a common practice to optimize LLM serving performance and resource use. DistServe~\cite{zhong2024distservedisaggregatingprefilldecoding} optimizes LLM serving by disaggregating prefill and decoding across GPUs, co-optimizing resource allocation and parallelism to boost throughput under latency constraints. Mooncake~\cite{qin2024mooncakekvcachecentricdisaggregatedarchitecture} employs a KVCache-centric disaggregated setup, separating prefill/decoding clusters, using underutilized GPU resources for KVCache, and adding a scheduler with overload rejection to enhance long-context throughput. MegaScale-Infer~\cite{zhu2025megascaleinferservingmixtureofexpertsscale} addresses MoE inference via disaggregating attention and FFN modules, enabling independent scaling, with ping-pong pipelining and a high-performance communication library to maximize GPU utilization and throughput. Finally, the SGLang team~\cite{lmsys2025deepseek} deploys DeepSeek with Prefill-Decode (PD) disaggregation and large-scale expert parallelism on 96 H100 GPUs leverages DeepEP, DeepGEMM, Two-batch Overlap (TBO), and Expert Parallelism Load Balancer (EPLB) to boost throughput, reduce costs, and balance workloads across GPUs. 

\begin{figure}[h]
\centering
\includegraphics[width=\columnwidth]{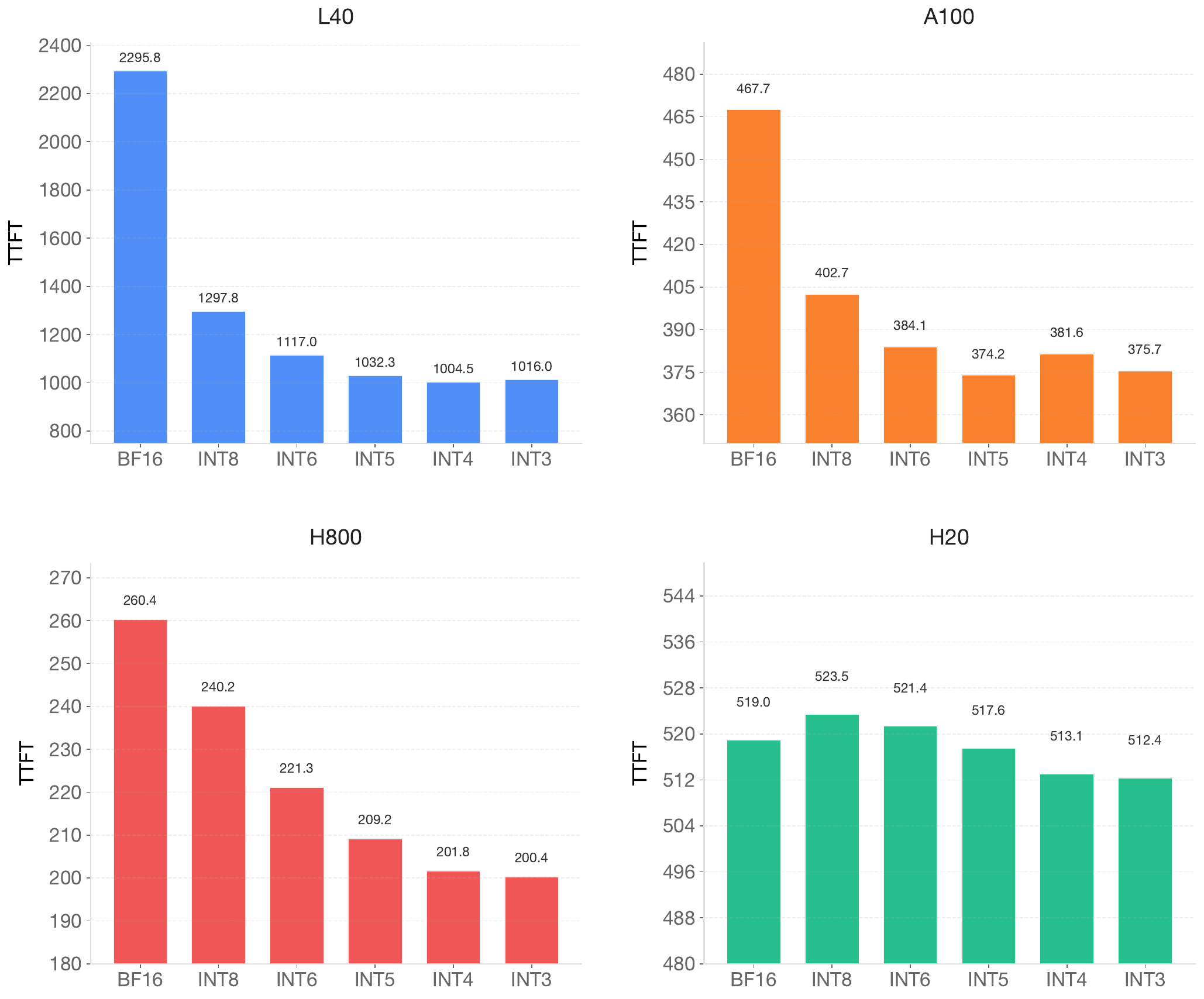} 
\caption{Llama-3-8B's TTFT across modern GPUs (TP=8) under various precision settings.}
\label{fig:ttft-any-bit-width}
\end{figure}

To resolve the widespread communication bottleneck introduced by distributed training (e.g., AllReduce in tensor parallelism, All2All communication in expert parallelism) and deployment (e.g., KV-cache communication in Prefill-decoding disaggregation), we introduce a novel communication optimization method called Flash Communication V2. Specifically, we compress the communication volume with quantization at any possible bit width, as shown in Figure~\ref{fig:c4-any-bit-width}. To achieve a practical performance boost, we employ \emph{bit-splitting}, which reorganizes the data storage mechanism to handle irregular bit widths. Also, we involve \emph{spike reserving} to manage outliers in lower precision without incurring too much overhead. Our overall results are demonstrated by Llama-3-8B's TTFT measurement on various modern GPUs in Figure~\ref{fig:ttft-any-bit-width}.

\section{Related Work}
%





\textbf{Communication and Computation Overlap.} Several works focus on optimizing communication and computation overlap in large-scale model training and inference. ScMoE~\cite{cai2025shortcutconnectedexpertparallelismaccelerating} addresses the All2All communication bottleneck by decoupling communication from conventional sequential ordering via a shortcut-connected MoE architecture, achieving 1.49$\times$ training and 1.82$\times$ inference speedups. Domino~\cite{wang2024dominoeliminatingcommunicationllm} eliminates AllReduce communication overhead in LLM training by breaking batch data dependency into micro-batches and pipelining them, yielding up to 1.3$\times$ speedup over Megatron-LM~\cite{MegatronLM}. DeepSeek-V2~\cite{liu2024deepseek} overlaps the computation of shared experts with the expert parallel All2All communication. DeepSeekV3~\cite{zhao2025insights} overlaps All2All communication of MoEs over attention computation through micro-batches scheduling, also adopted by MiniMax-01~\cite{minimax2025minimax01scalingfoundationmodels}. 

\textbf{Communication Quantization.} DeepSeekV3 quantizes activation into FP8 to have low-precision communication~\cite{deepseekai2025deepseekv3technicalreport}. ZeRO++~\cite{wang2023zero++} addresses the communication volume limitations of the Zero Redundancy Optimizer (ZeRO) in giant model training, especially in low-bandwidth clusters or at scale with small per-GPU batch sizes. It introduces three techniques: block-quantization based AllGather, data remapping that trades communication for memory, and an All2All-based quantized gradient averaging paradigm replacing reduce-scatter, collectively reducing ZeRO's communication volume by 4$\times$ and enabling up to 2.16$\times$ better throughput at 384 GPU scale. Flash Communication~\cite{li2024flashcommunicationreducingtensor}, on the other hand, targets the tensor-parallelism communication bottleneck in large language model inference, particularly on devices with limited bandwidth, using an optimized fusion kernel called Flash-AllReduce to boost intra-node communication speed by over 3$\times$ and reduce time-to-first-token by 2$\times$ with nearly no loss in model accuracy, as demonstrated across various up-to-date LLMs.

\textbf{Low-bit Quantization.} 
LLM quantization methods focus on reducing memory usage and boosting efficiency with minimal loss of accuracy, each with its unique approaches and trade-offs. AWQ~\cite{lin2024awq} protects 1\% activation-identified salient weight channels via up-scaling, generalizing well but focusing on weight-only quantization. OmniQuant~\cite{shao2023omniquant} optimizes quantization hyperparameters via learnable weight clipping and learnable equivalent transformation, supporting diverse settings but demanding more computation. For extremely low-bit quantization, Norm Tweaking~\cite{li2024norm} updates only the weights of normalization layers to mitigate the quantization errors, excelling in low-bit scenarios but relying on existing PTQ methods. QuIP~\cite{chee2023quip} uses incoherent matrices for 2-bit quantization with guarantees. EfficientQAT~\cite{chen2024efficientqat} makes QAT feasible for LLMs via block-wise and end-to-end training. Meanwhile, weight-activation quantization methods have been invented to maximize the utilization of low-bit computation on GPUs. As activation outliers are one of the main challenges in quantization, various methods focus on suppressing such outliers. QuaRot~\cite{ashkboos2024quarot} eliminates outliers via rotations for 4-bit inference, using random rotations that may not be optimal. SpinQuant~\cite{liu2024spinquant} employs learned rotations to outperform random ones, adding complexity. FlatQuant~\cite{sun2024flatquant} enhances flatness via learnable affine transformations, achieving state-of-the-art results but needing extra kernel optimization.

\section{Method}

Given prevalent communication bottlenecks in the distributed scenario, research interests have been attracted to reduce communication volumes by quantization and hiding communication cost behind computation. However, previous quantization methods are not directly applicable in the communication scenario. Besides, there is a lack of thorough investigation of any bitwidth compression and different communication mechanisms under NVLink-based and PCIe-based systems.

In this paper, we are motivated to further reduce the communication volume (typically activations) using low-bit quantization. Nevertheless, it is non-trivial to provide a solution. First, activations are found to carry numbers of massive magnitudes that can't be simply removed, which renders quantization particularly challenging~\cite{sun2024massive}. Second, low-bit quantization is hardly hardware-friendly, scarce works have been devoted to making it efficient and feasible in the real world. Third, we find that communication primitives are very sensitive to different quantization bitwidths (as later shown in Table~\ref{tab:ar-quant-c4-ppl-llama-qwen} and Table~\ref{tab:all2all-quant-moe-c4}). Therefore, we attempt to resolve these challenges and explore the possibility to achieve practical gains at various compression bitwidths.

\subsection{Quantization Sensitivity of Communication Primitives}

We first demonstrate the quantization sensitivity of two mainstream communication primitives using asymmetric fine-grained Round-to-Nearest quantization in Table~\ref{tab:ar-quant-c4-ppl-llama-qwen} and Table~\ref{tab:all2all-quant-moe-c4}. We follow Flash Communication~\cite{li2024flashcommunicationreducingtensor} for All-Reduce communication in TP since it substantially reduces QDQ steps compared with NCCL Ring-AllReduce. And we follow DeepSeek V3~\cite{deepseekai2025deepseekv3technicalreport} which only quantizes the dispatch volume in All2All communication for expert parallelism.

\begin{table}[ht]
\setlength{\tabcolsep}{3pt}
\centering
\begin{tabular}{lH*{6}{r}}
\toprule
Model	&	Method	&	INT8	&	INT6	&	INT5	&	INT4	&	INT3	&	INT2\\
\midrule
Llama-3-8B	&	Naive	&	8.89	&	8.94	&	9.07	&	9.67	&	13.72	&	7e5\\
\midrule
Llama-3-70B	&	Naive	&	6.74	&	6.75	&	6.81	&	7.05	&	8.40	&	1e2\\
\midrule
Qwen-3-8B	&	Naive	&	13.30	&	13.33	&	13.42	&	13.81	&	16.04	&	3e2\\
\midrule
Qwen-3-32B	&	Naive	&	10.78	&	10.81	&	10.89	&	11.21	&	12.93	&	2e2\\
\bottomrule
\end{tabular}
\caption{C4 Perplexity on Llama-3 and Qwen-3 models at various communication quantization (AllReduce in TP) bitwidths using naive RTN quantization. The default group size is set to 128.}
\label{tab:ar-quant-c4-ppl-llama-qwen}
\end{table}

Surprisingly, we discover at lower bitwdiths like INT5, it enjoys similar accuracy as INT8 in both scenarios while directly reducing above 30\% communication volume. On the contrary, All2All communication quantization at extreme bits like 2-bit still has feasible loss, while in the AllReduce scheme, the performance tremendously collapses.

\begin{table}[ht]
\centering
\setlength{\tabcolsep}{1.5pt}
\begin{tabular}{lHH*{6}{c}HHH}
\toprule
Model	&	BF16	&	Quant Method	&	INT8	&	INT6	&	INT5	&	INT4	&	INT3	&	INT2	&	INT4$_{gs32}$	&	INT3$_{gs32}$	&	INT2$_{gs32}$\\
\midrule
Qwen3-30B-A3B	&	9.65	&	RTN	&	9.65	&	9.66	&	9.7	&	9.88	&	10.61	&	19.71	&	9.72	&	9.99	&	11.67\\
\midrule
Qwen1.5-MoE-A2.7B	&	9.3	&	RTN	&	9.3	&	9.31	&	9.35	&	9.5	&	10.62	&	30.54	&	9.39	&	9.72	&	12.3\\
\bottomrule
\end{tabular}
\caption{C4 Perplexity of MoE models at various communication quantization (All2All dispatch in EP) bitwidths. The default group size is set to 128.}
\label{tab:all2all-quant-moe-c4}
\end{table}

\subsection{Bit Splitting}

\begin{figure}[h]
\centering
\includegraphics[width=\columnwidth]{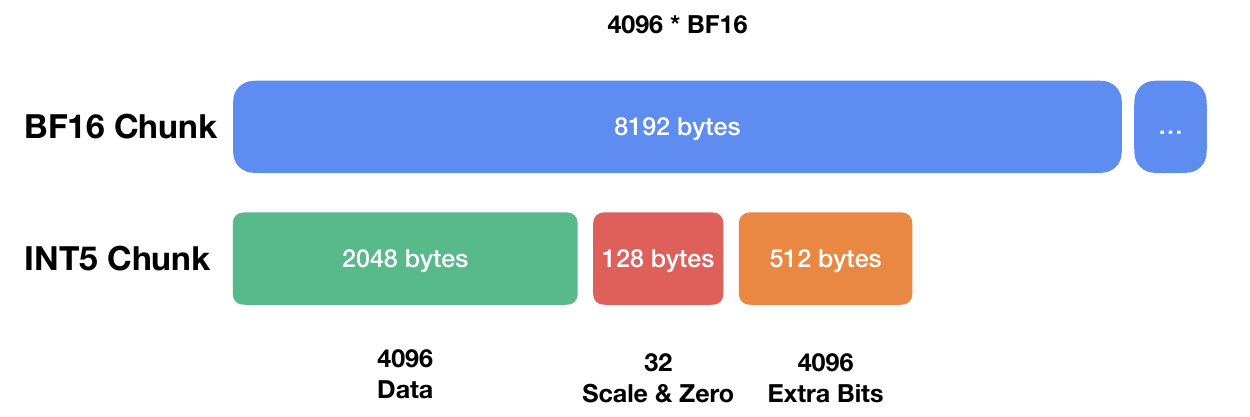}
\caption{Bit-splitting mechanism. We split the INT5 data into the first 4 bits and an extra singular bit. All 4-bit parts are saved together, so are the extra bits. Scales and zeros are in BF16.}
\label{fig:bit-split}
\end{figure}

Consequently, how to pack sub-bytes of such irregular bitwidth becomes critical for efficient transmission. To address this issue, we propose \emph{bit splitting}, which separates irregular bitwidths into regular parts (typically 4-bit or 2-bit) and a standalone extra bit. Regular parts of the same chunk are packed together while extra bits are independently stored, as exemplified in Figure~\ref{fig:bit-split}. For the regular part, we follow the fast packing strategy as in Flash Communication~\cite{li2024flashcommunicationreducingtensor}.

\subsection{Spike Reserving}

To alleviate the quantization loss of existing methods in extremely low-bit quantization, we attempt to increase the granularity of quantization by reducing the group size from 128 to 32. However, the simplest method round-to-nearest (RTN) still results in an unacceptable loss in the INT2 case, as shown in Table~\ref{tab:c4-ppl-llama-qwen-gs32}. We also apply advanced quantization techniques like Hadamard transformation~\cite{ashkboos2024quarot} and LogFMT~\cite{zhao2025insights}. Unfortunately, we observe that these methods behave even worse in extremely low-bit quantization, causing tremendous performance collapse in the INT2 scenario. Although it is effective to reduce the dynamic numerical range, the quantized range of Hadamard transformation and LogFMT is harder to recover since their dequantization enlarges the quantization errors (i.e., accumulative errors in Hadamard transformation and exponential amplification of errors in LogFMT). 

\begin{table}[t]
\centering
\begin{tabular}{l|c*{3}{|r}}
\toprule
Model	&	Method	&	INT4 	&	INT3	&	INT2 \\
\midrule
Llama-3-8B	&	RTN	&	9.2	&	10.54	&	40.59\\
	&	Hadamard	&	9.18	&	10.47	&	91.23\\
	&	LogFMT	&	9.3	&	11.53	&	1e3\\
	&	SpikeReserving	&	\textbf{9.01}	&	\textbf{9.57}	&	\textbf{14.39}\\
\midrule
Llama-3-70B	&	RTN	&	6.87	&	7.41	&	12.07\\
	&	Hadamard	&	6.97	&	9.22	&	5e3\\
	&	LogFMT	&	7.23	&	9.69	&	6e3\\
	&	SpikeReserving	&	\textbf{6.8}	&	\textbf{7.05}	&	\textbf{8.91}\\
\midrule
Qwen-3-8B	&	RTN	&	13.49	&	14.23	&	23.19\\
	&	Hadamard	&	13.51	&	14.16	&	26.63\\
	&	LogFMT	&	13.58	&	15.32	&	6e2\\
	&	SpikeReserving	&	\textbf{13.37}	&	\textbf{13.61}	&	\textbf{15.78}\\
\midrule
Qwen-3-32B	&	RTN	&	10.94	&	11.49	&	16.96\\
	&	Hadamard	&	10.94	&	11.61	&	30.17\\
	&	LogFMT	&	11.17	&	12.63	&	1e2\\
	&	SpikeReserving	&	\textbf{10.84}	&	\textbf{11.03}	&	\textbf{12.53}\\
\bottomrule
\end{tabular}
\caption{C4 Perplexity on Llama-3 and Qwen-3 models. For fine-grained quantization, the group size is set to 32, and asymmetric quantization is adopted for all bit-width settings.}
\label{tab:c4-ppl-llama-qwen-gs32}
\end{table}

Additionally, Hadamard transformation also introduces substantial computation overhead, which requires a couple of matrix multiplications for the transformation and recovering. LogFMT is also computationally expensive since non-linear transformations (e.g., log, exp) are costly operations in CUDA Math API~\footnote{\url{https://docs.nvidia.com/cuda/cuda-math-api/index.html}}. 

\begin{figure}[h]
\centering
\includegraphics[width=\columnwidth]{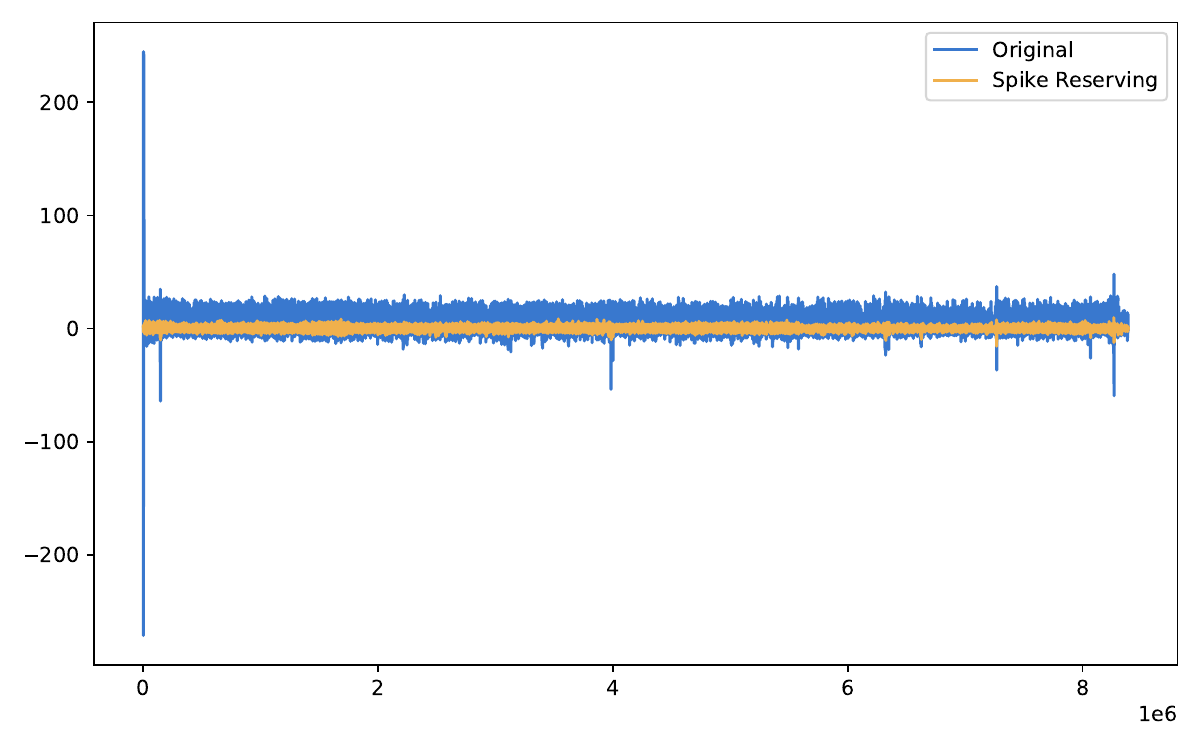}
\caption{Activation distribution of Llama-3-8B last layers' down$_{proj}$ before and after Spike Reserving.}
\label{fig:activation-quant}
\end{figure}

\begin{figure}[h]
\centering
\includegraphics[width=\columnwidth]{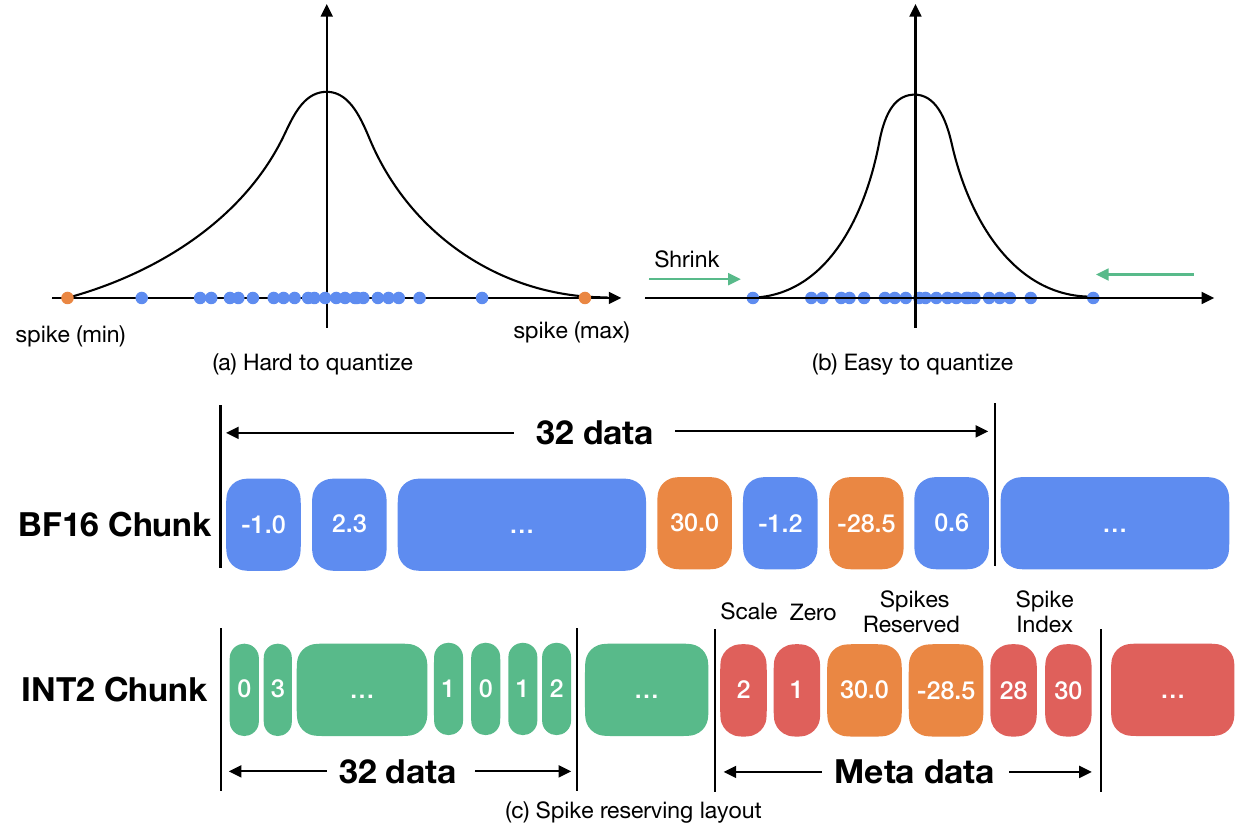}
\caption{Spike Reserving for INT2 compression. For a group of 32 numbers, we identify the spikes (min, max) and save them in the meta data section along with their indices.}
\label{fig:spike-reserve}
\end{figure}

Given the above challenges, we propose \emph{Spike Reserving}, which is simple to implement and is effective both for accuracy and efficiency. Observing the outliers in the low-bit scenario are typically located at the minima or the maxima point (we call them spikes, as in Figure~\ref{fig:spike-reserve} a), we separately store the minima of maxima of each quantization group in float precision, only to quantize the rest numbers in the shrinked numerical range, as exemplified in Figure~\ref{fig:spike-reserve} b. Take Llama-3-8B as an exmaple in Figure~\ref{fig:activation-quant}, when spikes are removed, the activation distribution becomes much easier to quantize as the numerical range is substantially narrowed. 

Figure~\ref{fig:spike-reserve} c shows the memory layout of spike reserving for INT2 compression. For each group of 32 numbers, we first identify the minimum and maximum and set them to zeros, the left data is quantized with the new minimum and maximum of the shrinked range. We include spikes and their indices in meta data section alongside scales and zeros. After dequantization, we restore the spikes to their original place. 

We also save scale as $scale_{int}$ as defined by,

\begin{equation}
    scale_{int} = \lfloor log_{2}{(scale)} * \theta \rceil
\end{equation}

where $\theta$ is typically set to 10 for linear upscaling. Spike indices are also saved as INT8 instead of BF16. Given 4096 float16 numbers that require 8192 bytes, we apply spike reserving to use only 2560 bytes.  With the integer scales and indices, it can further reduce the memory footprint by 20\%, as indicated by Table~\ref{tab:mem-footprint-saving}.

\begin{table}[ht]
\setlength{\tabcolsep}{2pt}
\centering
\begin{tabular}{*{7}{c}}
\toprule
Scheme & Data	&	Quantized	&	Scale\&zero	&	Spikes	&	Meta	&	Total$_{SR}$ \\
\midrule
$scale$& 8192	&	1024	&	512	&	1024	&	1536	&	2560 \\
$scale_{int}$& 8192	&	1024	&	256	&	768	&	1024	&	2048 \\
\bottomrule
\end{tabular}
\caption{Memory footprint saving (in bytes) of Spike Reserving with integer scales and indices in INT2 compression.}
\label{tab:mem-footprint-saving}
\end{table}

\subsection{Pipeline Parallelism in Hierarchical Communication}

Another dimension of efficient communication that we need to address is the challenge in low-bandwidth scenarios, such as popular cost-effective systems like the NVIDIA L20 and L40. For example, a node with 8$\times$L40 GPUs is not equipped with NVLink and has to use PCIe and NUMA bridges for communication. Hierarchical communication is the common practice in such a system ~\cite{jia2018highly,mikami2018massively,zhao2025insights}. 

\begin{figure}[h]
\centering
\includegraphics[width=\columnwidth]{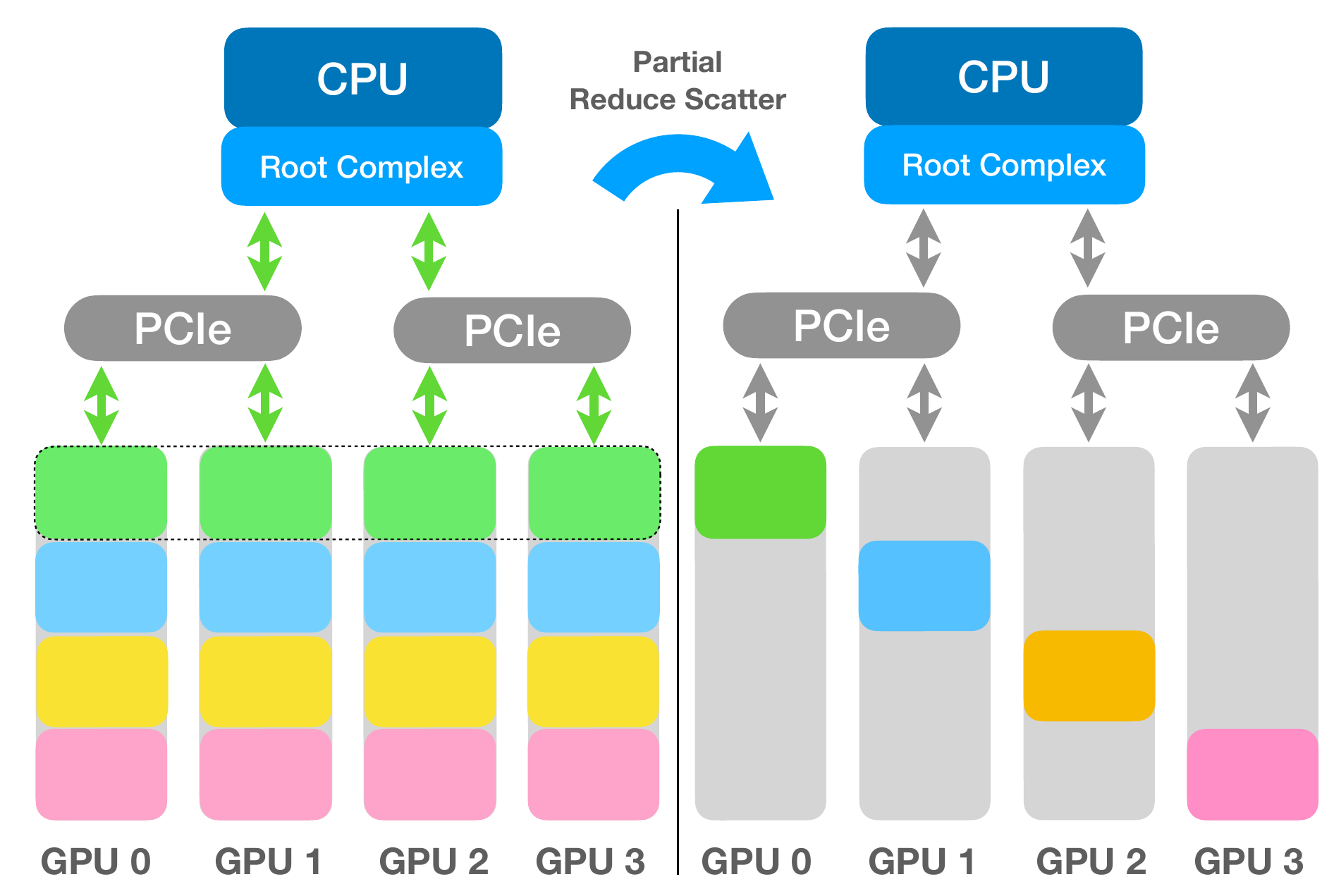}
\caption{Partial Reduce-Scatter is first performed within the same PCIe groups.}
\label{fig:hier-exec-reduce-scatter}
\end{figure}

\begin{figure}[h]
\centering
\includegraphics[width=\columnwidth]{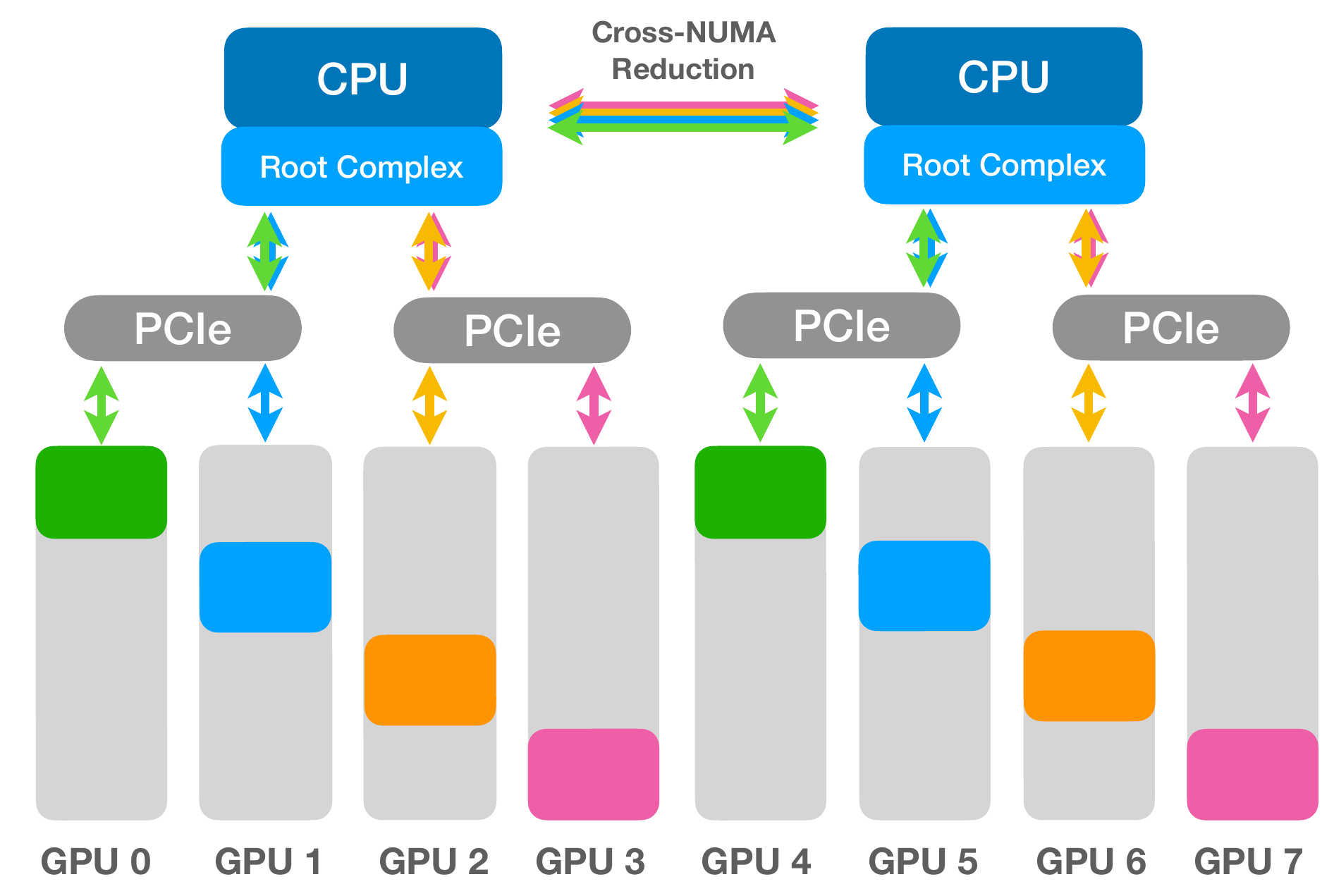}
\caption{Reduction is performed with cross-NUMA bridge after partial Reduce-Scatter.}
\label{fig:hier-exec-cross-numa}
\end{figure}

We also apply the hierarchical communication to the two-step scheme in Flash Communication~\cite{li2024flashcommunicationreducingtensor}. Take AllReduce communication for tensor parallelism as an example, we separate it into three stages including partial ReduceScatter within the same NUMA group, Reduction across NUMA bridges, and partial AllGather within the same NUMA group. Figure~\ref{fig:hier-exec-reduce-scatter} and ~\ref{fig:hier-exec-cross-numa} exhibit our hierarchical execution of three communication steps in an NVIDIA L40 GPU. For partial Reduce-Scatter operation, we first collect the data volume of the same NUMA group (e.g. GPU 0-3), the partial sum is then accumulated within each GPU after reduction. Next, only the partial sum goes through the NUMA bridge for the final reduction. The key difference lies in the amount of cross-NUMA data transferred during reduction: Two-step involves $4M$ total cross-NUMA data across its two steps ($M$ being the volume per GPU), while Hierarchical Two-step reduces this to M by limiting cross-NUMA data to $M/4$ per relevant GPU, with only 4 GPUs involved in point-to-point reduction, thus saving 3 times cross-NUMA communication volume (Table~\ref{tab:volume-comp}). 
We then perform AllGather within the same NUMA group to have the same data on each GPU (omitted for illustration).

\begin{table}[ht]
\setlength{\tabcolsep}{3pt}
\centering
\begin{tabular}{l*{2}{r}}
\toprule
Method	&	Volume$_{total}$	&	Volume$_{CrossNUMA}$ \\
\midrule
NCCL	&	14M	&	7M/4 \\
Two-step	&	14M	&	4M \\
Hierarchical Two-step	&	14M	&	M \\
\bottomrule
\end{tabular}
\caption{Volume comparison with Two-step, hierarchical Two-step and NCCL communication.}
\label{tab:volume-comp}
\end{table}

\begin{figure}[h]
\centering
\includegraphics[width=\columnwidth]{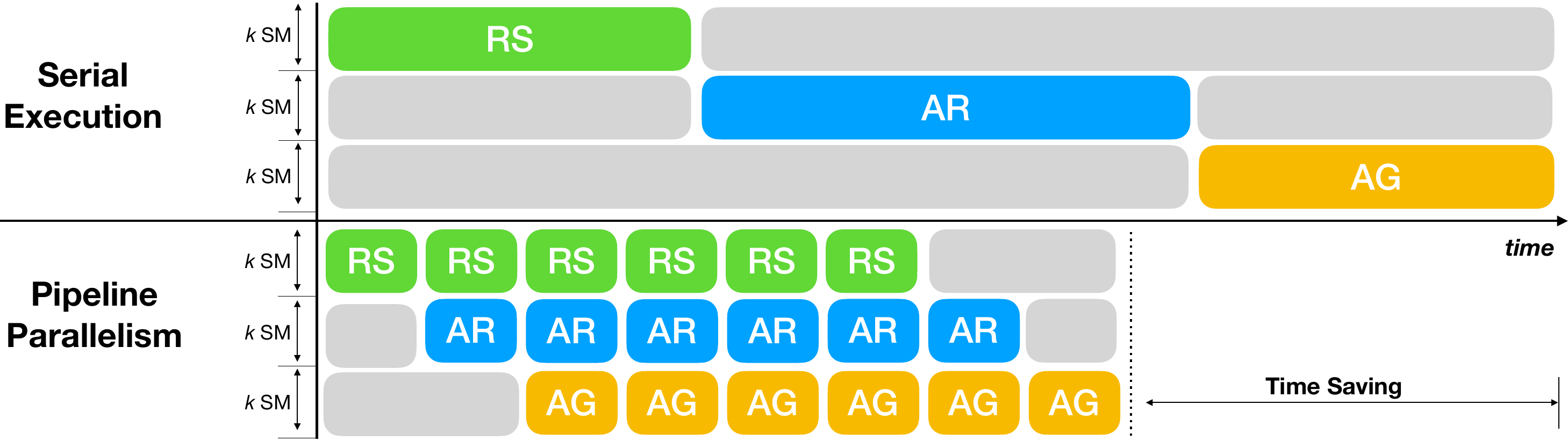}
\caption{Hierarchical Pipeline Parallelism vs. Serial Execution. RS: Reduce Scatter, AR: All Reduce, AG: All Gather. Gray areas indicate bubbles.}
\label{fig:hier-pp}
\end{figure}

However, during the serial execution of the hierarchical communication, the bandwidth is either idle or not utilized effectively (Figure~\ref{fig:hier-pp} top). Specifically, the NUMA bandwidth is idle during partial ReduceScatter while the PCIe bandwidth is under-utilized during cross-NUMA reduction. To maximize the bandwidth utilization, we involve pipeline parallelism as illustrated by Figure~\ref{fig:hier-pp} to split workloads into microchunks to create a pipeline, which is measured to have up to 20\% time saving.

\section{Experiments}

\subsection{Setup}
We measure the algorithmic communication bandwidths on common devices as stated in Table~\ref{tab:device-info}. We use LMDeploy~\cite{2023lmdeploy} for LLM inference. We focus on the communciation within a single node instead of a large cluster of GPUs.

\begin{table}[ht]
\setlength{\tabcolsep}{2pt}
\centering
\begin{tabular}{lrcrHr}
\toprule
GPU	&	SM	& Inter-Connect	& BW (GB/s)	&	BW$_{mem}$	&	BF16 (TFlops) \\
\midrule
L40	&	142	&	PCIe & 64	&	864	&	90.5  \\
A100	&	108	& NVLINK8	& 400	&	2TB/s	&	19.5 \\
H800	&	132	& NVLINK8	& 400	&	3.35TB/s	&	67 \\
H20	&	78	& NVLINK18	& 900	&	4TB/s	&	44  \\
\bottomrule
\end{tabular}
\caption{Common GPU inter-connection bandwidths, memory bandwidths, and CUDA core computation power in BF16. BW: Bandwidth}
\label{tab:device-info}
\end{table}

\begin{table*}[t]
\centering
\begin{tabular}{l*{2}{|c}*{7}{|r}}
\toprule
Model	&	Model BitW	&	Comm BitW	&	Group	&	PIQA	&	ARC$_C$	&	ARC$_E$	&	HS	&	WG	&	Avg\\
\midrule
Llama3-8B	&	FP16	&	FP16	&		&	80.79	&	53.41	&	77.69	&	79.16	&	72.77	&	72.76 \\
	&		&	INT8	&	128	&	80.58	&	52.47	&	77.40	&	79.09	&	73.09	&	72.53\\
	&		&	INT6	&	128	&	79.98	&	51.37	&	77.65	&	78.73	&	73.16	&	72.18\\
	&		&	INT5	&	128	&	79.98	&	50.51	&	77.40	&	78.87	&	72.22	&	71.80\\
	&		&	INT4	&	128	&	80.09	&	51.02	&	75.84	&	78.11	&	70.48	&	71.11\\
	&		&	INT3	&	32	&	73.50	&	40.10	&	64.31	&	71.58	&	63.30	&	62.56\\
	&		&	INT3$_{SR}$	&	32	&	76.93	&	46.08	&	71.55	&	75.22	&	68.82	&	\textbf{67.72}\\
	&		&	INT2	&	32	&	61.86	&	27.05	&	42.76	&	51.70	&	53.12	&	47.30\\
	&		&	INT2$_{SR}$	&	32	&	73.72	&	40.70	&	63.51	&	69.82	&	61.25	&	\textbf{61.80}\\
\midrule
Llama3-70B	&	FP16	&	FP16	&		&	84.55	&	64.33	&	85.86	&	84.89	&	80.35	&	80.00\\
	&		&	INT8	&	128	&	84.55	&	63.91	&	85.82	&	84.90	&	80.82	&	80.00\\
	&		&	INT6	&	128	&	84.11	&	61.69	&	85.44	&	84.87	&	80.35	&	79.29\\
	&		&	INT5	&	128	&	83.79	&	62.71	&	85.35	&	84.92	&	80.66	&	79.49\\
	&		&	INT4	&	128	&	83.13	&	61.35	&	83.33	&	84.69	&	78.93	&	78.29\\
	&		&	INT3	&	32	&	80.47	&	56.31	&	79.00	&	82.03	&	73.48	&	74.26\\
	&		&	INT3$_{SR}$	&	32	&	82.48	&	59.64	&	82.53	&	83.58	&	77.82	&	\textbf{77.21}\\
	&		&	INT2	&	32	&	75.30	&	47.53	&	68.43	&	75.81	&	61.48	&	65.71\\
	&		&	INT2$_{SR}$	&	32	&	80.79	&	55.97	&	78.49	&	81.18	&	74.03	&	\textbf{74.09}\\
\midrule
Qwen3-8B	&	FP16	&	FP16	&		&	77.26	&	56.74	&	80.81	&	75.05	&	67.96	&	71.56\\
	&		&	INT8	&	128	&	77.58	&	56.57	&	80.89	&	74.92	&	68.11	&	71.61\\
	&		&	INT6	&	128	&	77.69	&	56.83	&	80.81	&	74.89	&	67.64	&	71.57\\
	&		&	INT5	&	128	&	78.02	&	56.48	&	80.13	&	74.44	&	68.27	&	71.47\\
	&		&	INT4	&	128	&	76.82	&	55.63	&	79.38	&	74.16	&	67.72	&	70.74\\
	&		&	INT3	&	32	&	76.17	&	53.07	&	78.45	&	73.00	&	64.56	&	69.05\\
	&		&	INT3$_{SR}$	&	32	&	76.06	&	54.44	&	79.17	&	73.52	&	66.46	&	\textbf{69.93}\\
	&		&	INT2	&	32	&	67.85	&	36.95	&	58.54	&	57.09	&	55.96	&	55.28\\
	&		&	INT2$_{SR}$	&	32	&	71.27	&	46.42	&	69.23	&	67.23	&	60.46	&	\textbf{62.92}\\
\midrule
Qwen3-32B	&	FP16	&	FP16	&		&	81.88	&	60.75	&	83.12	&	82.53	&	73.16	&	76.29\\
	&		&	INT8	&	128	&	81.83	&	60.84	&	83.29	&	82.65	&	73.24	&	76.37\\
	&		&	INT6	&	128	&	81.66	&	60.92	&	83.38	&	82.62	&	71.74	&	76.06\\
	&		&	INT5	&	128	&	82.21	&	59.39	&	83.00	&	82.51	&	72.14	&	75.85\\
	&		&	INT4	&	128	&	80.58	&	61.69	&	82.79	&	82.09	&	70.64	&	75.56\\
	&		&	INT3	&	32	&	80.69	&	59.56	&	81.23	&	81.41	&	70.72	&	74.72\\
	&		&	INT3$_{SR}$	&	32	&	80.79	&	60.24	&	83.00	&	81.95	&	70.64	&	\textbf{75.32}\\
	&		&	INT2	&	32	&	73.56	&	47.01	&	68.35	&	73.02	&	62.75	&	64.94 \\
	&		&	INT2$_{SR}$	&	32	&	77.37	&	53.75	&	76.18	&	78.52	&	65.04	&	\textbf{70.17} \\
\bottomrule
\end{tabular}
\caption{Accuracy of Llama and Qwen models with various communication quantization strategies. SR: SpikeReserving}
\label{tab:lmeval}
\end{table*}

For efficient communication, we design a fusion kernel to performance quantization and communication, following Flash Communication~\cite{li2024flashcommunicationreducingtensor}. We use 48 SMs for data transmission on L40, A100 and H800, except for H20 we utilize all 78 SMs since it has the limited compute capacity. Each chunk of 4096 BF16 numbers are processed by a CUDA block. Each CUDA block includes 512 threads, and each thread processes 8 BF16 numbers. By default, we set the group size of fine-grained quantization to 128 for INT8/6/5, and 32 for INT4/3/2, where INT2 is enabled with spike reserving. For fast meta data access, we utilize the first four warps from the same CUDA block to access meta data in a vectorized manner.

\begin{table*}[t]
\centering
\setlength{\tabcolsep}{2pt}
\begin{tabular}{l*{11}{|c}}
\toprule
Model	&	BF16	&	Quant Method	&	INT8	&	INT6	&	INT5	&	INT4	&	INT3	&	INT2	&	INT4$_{gs32}$	&	INT3$_{gs32}$	&	INT2$_{gs32}$\\
\midrule
Qwen3-30B-A3B	&	9.65	&	RTN	&	9.65	&	9.66	&	9.7	&	9.88	&	10.61	&	19.71	&	9.72	&	9.99	&	11.67\\
	&		&	SR	&	9.65	&	9.65	&	9.66	&	9.71	&	9.92	&	11.55	&	9.67	&	9.75	&	10.3\\
\midrule
Qwen1.5-MoE-A2.7B	&	9.3	&	RTN	&	9.3	&	9.31	&	9.35	&	9.5	&	10.62	&	30.54	&	9.39	&	9.72	&	12.3\\
	&		&	SR	&	9.3	&	9.3	&	9.32	&	9.38	&	9.68	&	12.21	&	9.33	&	9.46	&	10.37 \\
\bottomrule
\end{tabular}
\caption{C4 Perplexity comparison between RTN and Spike Reserving. The default group size is set to 128, while $_{gs32}$ indicates a group size of 32. RTN: Round-To-Nearest, SR: SpikeReserving}
\label{tab:c4-moe-ppl}
\end{table*}

\begin{table*}[t]
\centering
\begin{tabular}{l*{7}{|r}}
\toprule
GPU	&	BF16$_{NCCL}$ &	INT8	&	INT6	&	INT5	&	INT4	&	INT3	&	INT2$_{SR}$ \\
\midrule
L40 (Two-step)	&	10.43	&	9.17	&	12.05	&	14.28	&	15.00	&	16.13	&	\textbf{16.19} \\
L40 (Hier) & - & 14.95	&	19.53	&	23.05	&	24.22	&	\textbf{28.80}	&	27.01 \\
L40 (HierPP)	&	- 	&	19.44	&	25.24	&	29.75	&	30.84	&	30.24	&	\textbf{33.39} \\
A100	&	89.15	&	123.32	&	137.72	&	147.68	&	148.61	&	\textbf{153.38}	&	122.46\\
H800	&	94.18	&	128.78	&	150.21	&	173.38	&	185.27	&	\textbf{187.31}	&	151.97\\
H20	&	209.14	&	214.32	&	225.55	&	240.59	&	\textbf{263.62}	&	260.35	&	202.11\\
\bottomrule
\end{tabular}
\caption{All-Reduce algorithmic bandwidths (in GB/s) on modern GPU devices under various communication schemes. SR: SpikeReserving. Hier: Hierarchical. HierPP: Hierarchical Pipeline Parallelism.}
\label{tab:all-reduce-bandwidth}
\vskip -0.15in
\end{table*}

\subsection{Quantization Accuracy}
We use LMEval~\cite{eval-harness} to evaluate the accuracy of dense models LLaMA-3 and Qwen-3 on PIQA~\cite{bisk2020piqa}, ARC\textsubscript{C} and ARC\textsubscript{E}~\cite{clark2018think}, HellaSwag \cite{zellers2019hellaswag}, WinoGrande \cite{sakaguchi2021winogrande} in various All-Reduce communication quantization bit widths, shown in Table~\ref{tab:lmeval}. Notice INT6/5 retain the accuracies close to INT8, which calls for an efficient implementation to benefit from communication reduction of lower-bitwidth. For the extremely low-bit quantization of INT3 and INT2, applying spike reserving gives up to \textbf{+14.5\%} accuracy boost.

Besides, we exhibit in Table~\ref{tab:c4-moe-ppl} the C4~\cite{C4} perplexity of All2All communication quantization in MoE models Qwen3-30B-A3B and Qwen1.5-MoE-A2.7B. Applying spike reserving in general delivers better performance than RTN. For extremely lower-bits, we can use the finer granularity $gs32$ to improve the performance.

\subsection{All-Reduce Communication}

We present the difference of All-Reduce communication (for expert parallelism) between naive RTN and our proposed Spike Reserving scheme in Table~\ref{tab:all-reduce-bandwidth}. Notice on L40, the two-step cross-NUMA communication volume is nearly twice the size of BF16, which cancels out the advantage of INT8 quantization, causing an algorithmic bandwidth of 9.17 GB/s (slightly lower than 10.43). But in all other low-bit cases, the quantization delivers observable gains. When applied with hierarchical communication and low-bit quantization, the advantage becomes more obvious, which is up to 3.2$\times$ compared with NCCL BF16.

Even in the high-bandwidth NVLink-interconnected scenario, low-bit quantization is beneficial to have up to 1.72$\times$, 1.99$\times$, 1.26$\times$, respectively on A100, H800, and H20. Note H800 has more CUDA core compute capacity so that its performance boost is higher than A100. Since H20 is low in compute capacity but high in communication bandwidth, it has the least gain. Interestingly, INT2 is not the most beneficial in such a high-bandwidth scenario, as the costs associated with spike reserving and QDQ end up negating the benefits derived from communication.

\subsection{All2All Communication}
We present the difference of All2All communication (for expert parallelism) between naive RTN and our proposed Spike Reserving scheme in Table~\ref{tab:all2all-bandwidth}. As a result, INT4 has the best algorithmic bandwidth, being 1.32$\times$ and 2.01$\times$ higher than NCCL BF16. Unfortunately, there is no benefit in the high-bandwidth system as H20. Note the measurement is based on naive All2All communication primitives, where each GPU is dispatched with the same amount of data. For advanced implementation like DeepEP~\cite{deepep2025}, we leave it for future investigation.

\begin{table}[ht]
\setlength{\tabcolsep}{1.5pt}
\centering
\begin{tabular}{l*{7}{r}}
\toprule
GPU	&	BF16	&	INT8	&	INT6	&	INT5	&	INT4	&	INT3	&	INT2$_{SR}$\\
\midrule
A100	&	165.41	&	183.99	&	208.77	&	212.15	&	\textbf{218.10}	&	206.22	&	177.02\\
H800	&	169.76	&	230.51	&	276.82	&	300.20	&	\textbf{341.87}	&	290.50	&	259.61\\
H20	&	\textbf{349.34}	&	323.52	&	276.73	&	289.06	&	339.00	&	290.33	&	249.53\\
\bottomrule
\end{tabular}
\caption{All2All algorithmic bandwidths (in GB/s) on modern GPU devices under various communication schemes. SR: Spike-Reserving. BF16 measured in NCCL.}
\label{tab:all2all-bandwidth}
\vskip -0.1in
\end{table}

\subsection{Time-To-First-Token in Dense Models}

As illustrated in Figure~\ref{fig:ttft-any-bit-width}, we observe the substantial TTFT gains, i.e., 2.28$\times$ in L40, 1.24$\times$ A100 and 1.3 $\times$ in H800, when applied with low-bit quantization with bit splitting. Note we also adopt pipeline parallelism for hierarchical communication on L40.  These observations are consistent with the results in algorithmic bandwidths. However, we don't find any benefit using low-bit quantization on H20.





\section{Conclusion}

In a nutshell, we present a practical communication optimization method called Flash Communication V2 to support efficient transmission at any-bit bitwidth. The key contribution comprises of a compact memory arrangement called bit splitting, simple but effective dynamic range shrinking technique named spike reserving. Additionally, we also address communication bottleneck by pipeline parallelism in hierarchical communication in PCIe-based systems like NVIDIA L40. Our new approach is thoroughly evaluated across the common dense and MoE models and language benchmarks, indicating a feasible solution to the deployment of large LLMs in distributed systems. Our work is limited in that there is a lack of evaluation on a large GPU cluster. In addition, we resort GPU compute resources for data compression. We suggest the communication module support more compute during the phase of hardware design.

\clearpage
\bibliography{aaai25}

\begin{thebibliography}{36}
\providecommand{\natexlab}[1]{#1}

\bibitem[{Ashkboos et~al.(2024)Ashkboos, Mohtashami, Croci, Li, Cameron, Jaggi,
  Alistarh, Hoefler, and Hensman}]{ashkboos2024quarot}
Ashkboos, S.; Mohtashami, A.; Croci, M.~L.; Li, B.; Cameron, P.; Jaggi, M.;
  Alistarh, D.; Hoefler, T.; and Hensman, J. 2024.
\newblock Quarot: Outlier-free 4-bit inference in rotated llms.
\newblock \emph{arXiv preprint arXiv:2404.00456}.

\bibitem[{Bisk et~al.(2020)Bisk, Zellers, Gao, Choi et~al.}]{bisk2020piqa}
Bisk, Y.; Zellers, R.; Gao, J.; Choi, Y.; et~al. 2020.
\newblock Piqa: Reasoning about physical commonsense in natural language.
\newblock In \emph{Proceedings of the AAAI conference on artificial
  intelligence}, volume~34, 7432--7439.

\bibitem[{Cai et~al.(2025)Cai, Jiang, Qin, Cui, Kim, and
  Huang}]{cai2025shortcutconnectedexpertparallelismaccelerating}
Cai, W.; Jiang, J.; Qin, L.; Cui, J.; Kim, S.; and Huang, J. 2025.
\newblock Shortcut-connected Expert Parallelism for Accelerating
  Mixture-of-Experts.
\newblock arXiv:2404.05019.

\bibitem[{Chee et~al.(2023)Chee, Cai, Kuleshov, and De~Sa}]{chee2023quip}
Chee, J.; Cai, Y.; Kuleshov, V.; and De~Sa, C.~M. 2023.
\newblock Quip: 2-bit quantization of large language models with guarantees.
\newblock \emph{Advances in Neural Information Processing Systems}, 36:
  4396--4429.

\bibitem[{Chen et~al.(2024)Chen, Shao, Xu, Wang, Gao, Zhang, and
  Luo}]{chen2024efficientqat}
Chen, M.; Shao, W.; Xu, P.; Wang, J.; Gao, P.; Zhang, K.; and Luo, P. 2024.
\newblock Efficientqat: Efficient quantization-aware training for large
  language models.
\newblock \emph{arXiv preprint arXiv:2407.11062}.

\bibitem[{Clark et~al.(2018)Clark, Cowhey, Etzioni, Khot, Sabharwal, Schoenick,
  and Tafjord}]{clark2018think}
Clark, P.; Cowhey, I.; Etzioni, O.; Khot, T.; Sabharwal, A.; Schoenick, C.; and
  Tafjord, O. 2018.
\newblock Think you have solved question answering? try arc, the ai2 reasoning
  challenge.
\newblock \emph{arXiv preprint arXiv:1803.05457}.

\bibitem[{DeepSeek-AI(2025)}]{deepseekai2025deepseekv3technicalreport}
DeepSeek-AI. 2025.
\newblock DeepSeek-V3 Technical Report.
\newblock arXiv:2412.19437.

\bibitem[{Gao et~al.(2021)Gao, Tow, Biderman, Black, DiPofi, Foster, Golding,
  Hsu, McDonell, Muennighoff, Phang, Reynolds, Tang, Thite, Wang, Wang, and
  Zou}]{eval-harness}
Gao, L.; Tow, J.; Biderman, S.; Black, S.; DiPofi, A.; Foster, C.; Golding, L.;
  Hsu, J.; McDonell, K.; Muennighoff, N.; Phang, J.; Reynolds, L.; Tang, E.;
  Thite, A.; Wang, B.; Wang, K.; and Zou, A. 2021.
\newblock A framework for few-shot language model evaluation.

\bibitem[{Jia et~al.(2018)Jia, Song, He, Wang, Rong, Zhou, Xie, Guo, Yang, Yu
  et~al.}]{jia2018highly}
Jia, X.; Song, S.; He, W.; Wang, Y.; Rong, H.; Zhou, F.; Xie, L.; Guo, Z.;
  Yang, Y.; Yu, L.; et~al. 2018.
\newblock Highly scalable deep learning training system with mixed-precision:
  Training imagenet in four minutes.
\newblock \emph{arXiv preprint arXiv:1807.11205}.

\bibitem[{Kimi(2025)}]{kimik2technicalreport}
Kimi. 2025.
\newblock KIMI K2: OPEN AGENTIC INTELLIGENCE.

\bibitem[{Li et~al.(2024{\natexlab{a}})Li, Li, Zhang, and Chu}]{li2024norm}
Li, L.; Li, Q.; Zhang, B.; and Chu, X. 2024{\natexlab{a}}.
\newblock Norm tweaking: High-performance low-bit quantization of large
  language models.
\newblock In \emph{Proceedings of the AAAI Conference on Artificial
  Intelligence}, volume~38, 18536--18544.

\bibitem[{Li et~al.(2024{\natexlab{b}})Li, Zhang, Ye, Zhang, Wu, Sun, Ma, and
  Xie}]{li2024flashcommunicationreducingtensor}
Li, Q.; Zhang, B.; Ye, L.; Zhang, Y.; Wu, W.; Sun, Y.; Ma, L.; and Xie, Y.
  2024{\natexlab{b}}.
\newblock Flash Communication: Reducing Tensor Parallelization Bottleneck for
  Fast Large Language Model Inference.
\newblock arXiv:2412.04964.

\bibitem[{Lin et~al.(2024)Lin, Tang, Tang, Yang, Chen, Wang, Xiao, Dang, Gan,
  and Han}]{lin2024awq}
Lin, J.; Tang, J.; Tang, H.; Yang, S.; Chen, W.-M.; Wang, W.-C.; Xiao, G.;
  Dang, X.; Gan, C.; and Han, S. 2024.
\newblock Awq: Activation-aware weight quantization for on-device llm
  compression and acceleration.
\newblock \emph{Proceedings of machine learning and systems}, 6: 87--100.

\bibitem[{Liu et~al.(2024{\natexlab{a}})Liu, Feng, Wang, Wang, Liu, Zhao,
  Dengr, Ruan, Dai, Guo et~al.}]{liu2024deepseek}
Liu, A.; Feng, B.; Wang, B.; Wang, B.; Liu, B.; Zhao, C.; Dengr, C.; Ruan, C.;
  Dai, D.; Guo, D.; et~al. 2024{\natexlab{a}}.
\newblock Deepseek-v2: A strong, economical, and efficient mixture-of-experts
  language model.
\newblock \emph{arXiv preprint arXiv:2405.04434}.

\bibitem[{Liu et~al.(2024{\natexlab{b}})Liu, Zhao, Fedorov, Soran, Choudhary,
  Krishnamoorthi, Chandra, Tian, and Blankevoort}]{liu2024spinquant}
Liu, Z.; Zhao, C.; Fedorov, I.; Soran, B.; Choudhary, D.; Krishnamoorthi, R.;
  Chandra, V.; Tian, Y.; and Blankevoort, T. 2024{\natexlab{b}}.
\newblock Spinquant: Llm quantization with learned rotations.
\newblock \emph{arXiv preprint arXiv:2405.16406}.

\bibitem[{LMDeploy(2023)}]{2023lmdeploy}
LMDeploy. 2023.
\newblock LMDeploy: A Toolkit for Compressing, Deploying, and Serving LLM.
\newblock \url{https://github.com/InternLM/lmdeploy}.

\bibitem[{Mikami et~al.(2018)Mikami, Suganuma, Tanaka, Kageyama
  et~al.}]{mikami2018massively}
Mikami, H.; Suganuma, H.; Tanaka, Y.; Kageyama, Y.; et~al. 2018.
\newblock Massively distributed SGD: ImageNet/ResNet-50 training in a flash.
\newblock \emph{arXiv preprint arXiv:1811.05233}.

\bibitem[{MiniMax(2025)}]{minimax2025minimax01scalingfoundationmodels}
MiniMax. 2025.
\newblock MiniMax-01: Scaling Foundation Models with Lightning Attention.
\newblock arXiv:2501.08313.

\bibitem[{{NVIDIA}({2024})}]{MegatronLM}
{NVIDIA}. {2024}.
\newblock {Megatron-LM: Ongoing research training transformer models at scale}.
\newblock {GitHub repository}.

\bibitem[{Qin et~al.(2024)Qin, Li, He, Zhang, Wu, Zheng, and
  Xu}]{qin2024mooncakekvcachecentricdisaggregatedarchitecture}
Qin, R.; Li, Z.; He, W.; Zhang, M.; Wu, Y.; Zheng, W.; and Xu, X. 2024.
\newblock Mooncake: A KVCache-centric Disaggregated Architecture for LLM
  Serving.
\newblock arXiv:2407.00079.

\bibitem[{Qwen(2025)}]{yang2025qwen3technicalreport}
Qwen. 2025.
\newblock Qwen3 Technical Report.
\newblock arXiv:2505.09388.

\bibitem[{Raffel et~al.(2020)Raffel, Shazeer, Roberts, Lee, Narang, Matena,
  Zhou, Li, and Liu}]{C4}
Raffel, C.; Shazeer, N.; Roberts, A.; Lee, K.; Narang, S.; Matena, M.; Zhou,
  Y.; Li, W.; and Liu, P. 2020.
\newblock Exploring the Limits of Transfer Learning with a Unified Text-to-Text
  Transformer.
\newblock \emph{Journal of Machine Learning Research}, 21(140): 1--67.

\bibitem[{Sakaguchi et~al.(2021)Sakaguchi, Bras, Bhagavatula, and
  Choi}]{sakaguchi2021winogrande}
Sakaguchi, K.; Bras, R.~L.; Bhagavatula, C.; and Choi, Y. 2021.
\newblock Winogrande: An adversarial winograd schema challenge at scale.
\newblock \emph{Communications of the ACM}, 64(9): 99--106.

\bibitem[{Shao et~al.(2023)Shao, Chen, Zhang, Xu, Zhao, Li, Zhang, Gao, Qiao,
  and Luo}]{shao2023omniquant}
Shao, W.; Chen, M.; Zhang, Z.; Xu, P.; Zhao, L.; Li, Z.; Zhang, K.; Gao, P.;
  Qiao, Y.; and Luo, P. 2023.
\newblock Omniquant: Omnidirectionally calibrated quantization for large
  language models.
\newblock \emph{arXiv preprint arXiv:2308.13137}.

\bibitem[{StepFun(2025)}]{stepfun2025step3largeaffordablemodelsystem}
StepFun. 2025.
\newblock Step-3 is Large yet Affordable: Model-system Co-design for
  Cost-effective Decoding.
\newblock arXiv:2507.19427.

\bibitem[{Sun et~al.(2024{\natexlab{a}})Sun, Chen, Kolter, and
  Liu}]{sun2024massive}
Sun, M.; Chen, X.; Kolter, J.~Z.; and Liu, Z. 2024{\natexlab{a}}.
\newblock Massive activations in large language models.
\newblock \emph{arXiv preprint arXiv:2402.17762}.

\bibitem[{Sun et~al.(2024{\natexlab{b}})Sun, Liu, Bai, Bao, Zhao, Li, Hu, Yu,
  Hou, Yuan et~al.}]{sun2024flatquant}
Sun, Y.; Liu, R.; Bai, H.; Bao, H.; Zhao, K.; Li, Y.; Hu, J.; Yu, X.; Hou, L.;
  Yuan, C.; et~al. 2024{\natexlab{b}}.
\newblock Flatquant: Flatness matters for llm quantization.
\newblock \emph{arXiv preprint arXiv:2410.09426}.

\bibitem[{Team and Contributors(2025)}]{lmsys2025deepseek}
Team, L.; and Contributors. 2025.
\newblock Deploying DeepSeek with PD Disaggregation and Large-Scale Expert
  Parallelism on 96 H100 GPUs.
\newblock \url{https://lmsys.org/blog/2025-05-05-large-scale-ep/}.
\newblock Accessed: 2025-05-05.

\bibitem[{Wang et~al.(2023)Wang, Qin, Jacobs, Holmes, Rajbhandari, Ruwase, Yan,
  Yang, and He}]{wang2023zero++}
Wang, G.; Qin, H.; Jacobs, S.~A.; Holmes, C.; Rajbhandari, S.; Ruwase, O.; Yan,
  F.; Yang, L.; and He, Y. 2023.
\newblock Zero++: Extremely efficient collective communication for giant model
  training.
\newblock \emph{arXiv preprint arXiv:2306.10209}.

\bibitem[{Wang et~al.(2024)Wang, Zhang, Shen, Li, and
  Ruwase}]{wang2024dominoeliminatingcommunicationllm}
Wang, G.; Zhang, C.; Shen, Z.; Li, A.; and Ruwase, O. 2024.
\newblock Domino: Eliminating Communication in LLM Training via Generic Tensor
  Slicing and Overlapping.
\newblock arXiv:2409.15241.

\bibitem[{Zellers et~al.(2019)Zellers, Holtzman, Bisk, Farhadi, and
  Choi}]{zellers2019hellaswag}
Zellers, R.; Holtzman, A.; Bisk, Y.; Farhadi, A.; and Choi, Y. 2019.
\newblock Hellaswag: Can a machine really finish your sentence?
\newblock \emph{arXiv preprint arXiv:1905.07830}.

\bibitem[{Zhao et~al.(2025{\natexlab{a}})Zhao, Deng, Ruan, Dai, Gao, Li, Zhang,
  Huang, Zhou, Ma et~al.}]{zhao2025insights}
Zhao, C.; Deng, C.; Ruan, C.; Dai, D.; Gao, H.; Li, J.; Zhang, L.; Huang, P.;
  Zhou, S.; Ma, S.; et~al. 2025{\natexlab{a}}.
\newblock Insights into deepseek-v3: Scaling challenges and reflections on
  hardware for ai architectures.
\newblock In \emph{Proceedings of the 52nd Annual International Symposium on
  Computer Architecture}, 1731--1745.

\bibitem[{Zhao et~al.(2025{\natexlab{b}})Zhao, Zhou, Zhang, Deng, Xu, Liu, Yu,
  Li, and Zhao}]{deepep2025}
Zhao, C.; Zhou, S.; Zhang, L.; Deng, C.; Xu, Z.; Liu, Y.; Yu, K.; Li, J.; and
  Zhao, L. 2025{\natexlab{b}}.
\newblock DeepEP: an efficient expert-parallel communication library.
\newblock \url{https://github.com/deepseek-ai/DeepEP}.

\bibitem[{Zhong et~al.(2024)Zhong, Liu, Chen, Hu, Zhu, Liu, Jin, and
  Zhang}]{zhong2024distservedisaggregatingprefilldecoding}
Zhong, Y.; Liu, S.; Chen, J.; Hu, J.; Zhu, Y.; Liu, X.; Jin, X.; and Zhang, H.
  2024.
\newblock DistServe: Disaggregating Prefill and Decoding for Goodput-optimized
  Large Language Model Serving.
\newblock arXiv:2401.09670.

\bibitem[{Zhu et~al.(2025)Zhu, Jiang, Jin, Wu, Stuardo, Wang, Zhang, Zhou, Wei,
  Cheng, Xiao, Zhang, Liu, Lin, Chang, Ye, Yu, Liu, Jin, and
  Liu}]{zhu2025megascaleinferservingmixtureofexpertsscale}
Zhu, R.; Jiang, Z.; Jin, C.; Wu, P.; Stuardo, C.~A.; Wang, D.; Zhang, X.; Zhou,
  H.; Wei, H.; Cheng, Y.; Xiao, J.; Zhang, X.; Liu, L.; Lin, H.; Chang, L.-W.;
  Ye, J.; Yu, X.; Liu, X.; Jin, X.; and Liu, X. 2025.
\newblock MegaScale-Infer: Serving Mixture-of-Experts at Scale with
  Disaggregated Expert Parallelism.
\newblock arXiv:2504.02263.

\bibitem[{Zuo et~al.(2025)Zuo, Lin, Deng, Zou, Yang, Diao, Gao, Xu, Chen, Lu,
  Qiu, Li, Chang, Yu, Miao, Zheng, Li, Feng, Wang, Zong, Zhou, Zhou, Chen,
  Liao, Li, Zhang, Zhu, Wang, Xiao, Liang, Cao, Liu, Yang, Bai, Li, Xie, Wu,
  Yu, Chen, Liu, Ding, Zhu, Xia, Xiong, Yu, and
  Liao}]{zuo2025servinglargelanguagemodels}
Zuo, P.; Lin, H.; Deng, J.; Zou, N.; Yang, X.; Diao, Y.; Gao, W.; Xu, K.; Chen,
  Z.; Lu, S.; Qiu, Z.; Li, P.; Chang, X.; Yu, Z.; Miao, F.; Zheng, J.; Li, Y.;
  Feng, Y.; Wang, B.; Zong, Z.; Zhou, M.; Zhou, W.; Chen, H.; Liao, X.; Li, Y.;
  Zhang, W.; Zhu, P.; Wang, Y.; Xiao, C.; Liang, D.; Cao, D.; Liu, J.; Yang,
  Y.; Bai, X.; Li, Y.; Xie, H.; Wu, H.; Yu, Z.; Chen, L.; Liu, H.; Ding, Y.;
  Zhu, H.; Xia, J.; Xiong, Y.; Yu, Z.; and Liao, H. 2025.
\newblock Serving Large Language Models on Huawei CloudMatrix384.
\newblock arXiv:2506.12708.

\end{thebibliography}
\newpage

\end{document}